\documentclass[aps,pre,preprint,superscriptaddress]{revtex4-1}

\usepackage{graphicx} 
\usepackage{subfigure}
\usepackage{color}
\usepackage{comment}

\begin{document}


\title{Adaptive long-range migration promotes cooperation under tempting conditions}


\author{Genki Ichinose}
\email[Correspondence to: ]{ichinose@anan-nct.ac.jp}
\homepage[]{https://sites.google.com/site/igenki/}

\author{Masaya Saito}
\affiliation{Anan National College of Technology 265 Aoki Minobayashi, Anan, Tokushima 774-0017, Japan}

\author{Hiroki Sayama}
\affiliation{Collective Dynamics of Complex Systems Research Group, Binghamton University, State University of New York, Binghamton, NY 13902-6000, USA}

\author{David Sloan Wilson}
\affiliation{Departments of Biological Sciences and Anthropology, Binghamton University, State University of New York, Binghamton, NY 13902-6000, USA}


\date{\today}

\begin{abstract}
Migration is a fundamental trait in humans and animals. Recent studies investigated
the effect of migration on the evolution of cooperation, showing that contingent
migration favors cooperation in spatial structures. In those studies, only local migration to immediate neighbors
was considered, while long-range migration has not been considered yet, partly because the long-range migration has been generally regarded as harmful for cooperation as it would bring the population to a well-mixed state that favors defection.
Here, we studied the effects of adaptive long-range migration on the evolution of cooperation through agent-based 
simulations of a spatial Prisoner's Dilemma game where individuals can 
jump to a farther site if they are surrounded by more defectors.
Our results show that adaptive long-range migration strongly promotes cooperation,
especially under conditions where the temptation to defect is considerably high.
These findings demonstrate the significance of adaptive long-range migration for the evolution of cooperation.
\end{abstract}


\maketitle

\section*{Introduction}
Understanding the evolution of cooperation is one of the major
challenges in both social and biological sciences \cite{Axelrod1984}. 
Cooperators benefit others at the costs of their own, while defectors may enjoy benefits brought by others without paying any cost themselves.
It is well understood that, in a well-mixed population, cooperation cannot evolve because defectors are always better off than cooperators, unless special mechanisms are considered \cite{Nowak2006}.
In contrast, it is also known that spatial structures generally promote the evolution of cooperation
because spatial locality helps cooperators interact with other
cooperators more often than with defectors \cite{NowakMay1992}.
This is called network (or spatial) reciprocity \cite{Nowak2006}.
The evolution of cooperation has been considered on various types of networks \cite{SzaboFath2007, Wang_etal2012d, Wang_etal2013a}.
Furthermore, network reciprocity has been combined with punishment \cite{SzolnokiPerc2013} or wisdom of group \cite{Szolnoki_etal2012}, which are also known as promoting cooperative behavior.
The effect of reciprocity is more apparent when vacant sites are allowed in space \cite{MitteldorfWilson2000}.
In those situations, Wang et al.~\cite{Wang_etal2012a, Wang_etal2012b} found that cooperation was most promoted when the population density was close to the percolation threshold of underlying graph.

In a spatially extended setting, migration, a fundamental behavioral trait of animals and humans, can change
dynamics of evolutionary games.
Migration plays a key role in keeping biodiversity \cite{Reichenbach_etal2007, Jiang_etal2011, Jiang_etal2012} as well as the evolution of cooperation in the real world \cite{Kerr_etal2006}.
Researchers have thus considered the effect of migration in spatial structures on cooperative behavior.
Moreover, it has been suggested that coevolution of strategies and mobilities strongly enhances cooperation.
This is the important aspect of ``coevolutionary games" \cite{PercSzolnoki2010}.

Earlier studies on random migration have shown that high migration rates generally prevent the evolution of cooperation
since it brings the population closer to a well-mixed state, while cooperation may still be sustained if migration rates are moderate \cite{EnquistLeimar1993, RichersonBoyd2005, TraulsenNowak2006, Vainstein_etal2007}.
However, it has recently been shown that ``contingent migration" has a different effect on the evolution of cooperation.
In contingent migration models, each individual monitors its current environmental condition within its local neighborhood and migrates to another location if the condition is found to be undesirable.
For instance, Jiang et al.~\cite{Jiang_etal2010} studied a model of spatial games in which an individual moves to a nearby empty site with probability $n_D/4$, where $n_D$ is the number of defectors in four adjacent neighbor sites. In other words, the individuals change their migratory decisions adaptively in response to the level of defectors' presence. They showed that this type of ``adaptive migration'' promotes cooperation, especially with an intermediate population density. Similarly, Yang et al.~\cite{Yang_etal2010} proposed a model with aspiration-induced migration in which an individual moves to a randomly chosen empty site within its four neighbors if the payoff of the individual falls below its aspiration level. 
They found that the optimal level of cooperation was achieved when both the aspiration level and the population density were moderate.
Another movement from unfavorable locations was implemented by Chen et al.~\cite{Chen_etal2012}, where, the lower the contribution from individuals was, the more individuals migrated from each location.
Moreover, Helbing et al.~\cite{HelbingYu2008, Helbing2009, HelbingYu2009} showed that cooperation is even more facilitated if individuals can choose a preferred destination to move in the adaptive migration.
These earlier studies have collectively illustrated that such adaptive migration strongly promotes the evolution of cooperation,
allowing cooperative individuals to avoid exploitation by defectors.

One limitation in the earlier contingent migration studies is that
migration was commonly assumed to be limited only to immediate
neighbor sites of each individual. In the meantime, it has recently been shown that long-range migration
can also promote cooperation if the strength of between-group selection is increased \cite{FuNowak2013}.
To our knowledge of the relevant literature, no existing work addressed the effect of
{\em adaptive long-distance migration} on the evolution of cooperation, partly because the long-range migration has been generally regarded as harmful for cooperation as it would bring the population to a well-mixed state that favors defection.
We believe, however, that adaptive long-range migration may be beneficial for
cooperators because they may be able to escape from defector-dominated areas at
once and potentially land in a community of other cooperators.
This adaptive long-range migration is important because it is naturally observed in both humans and animals.

In this paper, we studied the effects of adaptive long-range migration on the evolution of cooperation. We develop an agent-based model in which individuals play
the Prisoner's Dilemma (PD) game with local nearest neighbors on a regular spatial grid and may migrate
to another site if the local situation is unfavorable. The key unique contribution of our work 
is to use an adaptive distance for long-range migration based on the number
of defectors in the individual's vicinity, so that the possible distance of the adaptive migration is longer with more
defectors in the area nearby.

\section*{Results}
\begin{figure*}[!t]
\centering
\includegraphics[width=\textwidth]{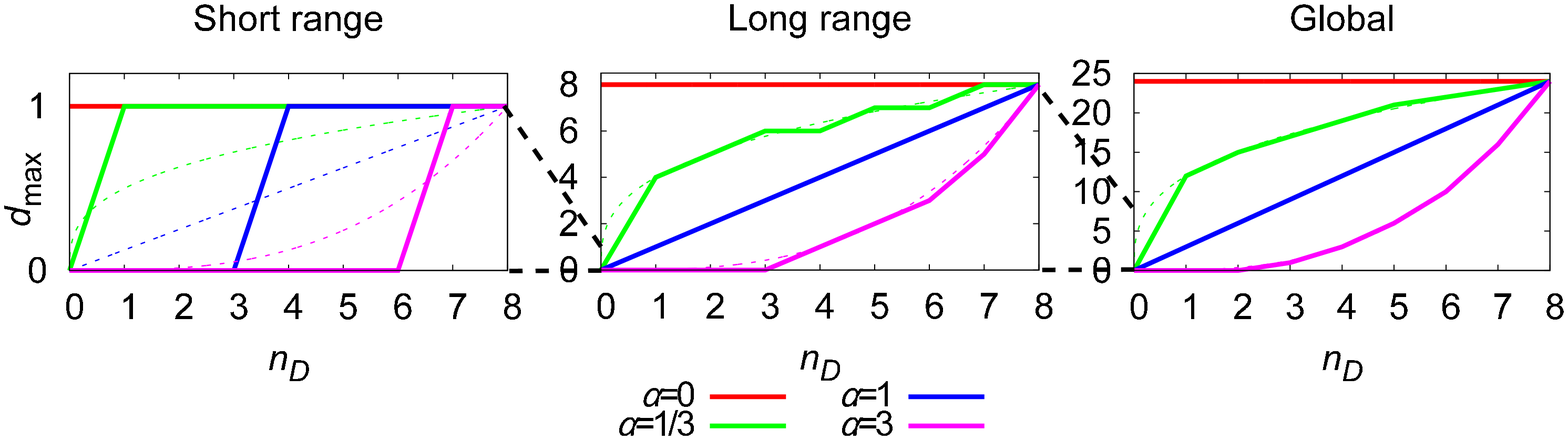}
\caption{Maximum migration distance $d_{\mathrm{max}}$ as a function of the number of defectors in the neighborhood ($n_D$). The Moore neighborhood is assumed (i.e., $n=8$). Left: Short-range migration ($d=1$). Center: Long-range migration ($d=8$). Right: Global migration ($d=24 \sim L/2$). Random migration models do not depend on $n_D$ (red lines at the top), while adaptive migration models (all the other curves) depend on $n_D$.}
\label{d_max}
\end{figure*}

\begin{figure}[!t]
\centering
\includegraphics[width=2.5in]{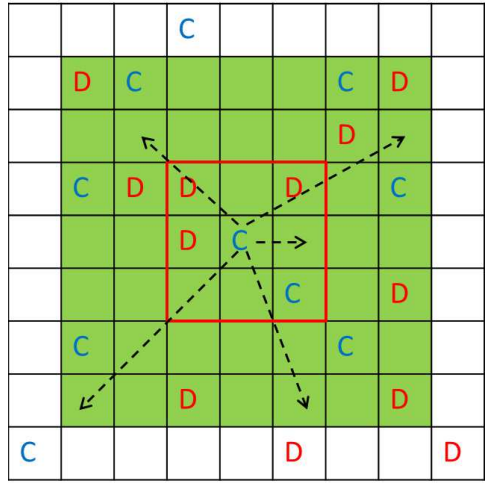}
\caption{An illustration of how ALM (Adaptive Long-range Migration) works ($\alpha=1$, $d=8$). In this example, there are three defectors within eight neighbors indicated by the red square (i.e., $n=8$, $n_D=3$). The maximum migration distance is calculated as follows: $d_{\mathrm{max}}=\mathrm{round}((3/8)^1 \times 8)=3$. Therefore, the focal cooperator can migrate to another empty site within the 7-by-7 local region indicated in green.}
\label{d_max_sample}
\end{figure}

We conducted simulation experiments using the agent-based model described in the \textbf{Methods}.
The parameter setting used for the experiments is as follows unless noted otherwise: $L=50$, $\rho=0.3$, $n=8$ (Moore neighborhood).
See the \textbf{Methods} section for the meaning of each parameter.
We systematically varied values of $\alpha$ (sensitivity to defectors; note that smaller values of $\alpha$ means high sensitivity, except for $\alpha = 0$) and $d$ (migration range) to represent the following six different types of migration:
\begin{itemize}
\item $\alpha=0$; $d=1$: random short-range migration (RSM)
\item $\alpha=0$; $d=8$ $(> 1)$: random long-range migration (RLM)
\item $\alpha=0$; $d=24$ $(\sim L/2)$: random global migration (RGM; an extreme case of RLM)
\item $\alpha=1/3, 1,$ or $3$; $d=1$: adaptive short-range migration (ASM)
\item $\alpha=1/3, 1,$ or $3$; $d=8$ $(> 1)$: adaptive long-range migration (ALM)
\item $\alpha=1/3, 1,$ or $3$; $d=24$ $(\sim L/2)$: adaptive global migration (AGM; an extreme case of ALM)
\end{itemize}
Figure \ref{d_max} shows $d_{\mathrm{max}}(n_D)$ in each case. $d_{\mathrm{max}}$ in random migration cases (denoted RXMs hereafter) do not depend on $n_D$, while $d_{\mathrm{max}}$ in adaptive migration cases (denoted AXMs hereafter) monotonically increases along $n_D$.
RSM is a form of random migration to the immediate neighbors, which corresponds to typical assumptions adopted in earlier models. ASM corresponds to adaptive migration used in the recent literature.
Figure \ref{d_max_sample} illustrates an example of ALM where the focal individual (located at the center) counts the number of defectors within the neighbors and tries to move to an empty site within a $d_{\mathrm{max}}$ range.

Other experimental parameters being varied included $T$, the temptation to defect, and $\rho$, the population density, among others.
The main experimental measurement obtained from each simulation run is the equilibrium fraction of cooperators, $\bar{f}_c$, after a certain time period elapses. This is obtained by averaging $f_c(t)$, the fraction of cooperators at generation $t$, for $t=4,001\sim5,000$.
The results are also averaged over 20 independent simulation runs for each set of parameters to check the robustness.

\subsection*{Effect of adaptive long-range migration under tempting conditions}
\begin{figure}[!t]
\centering
\includegraphics[width=5in]{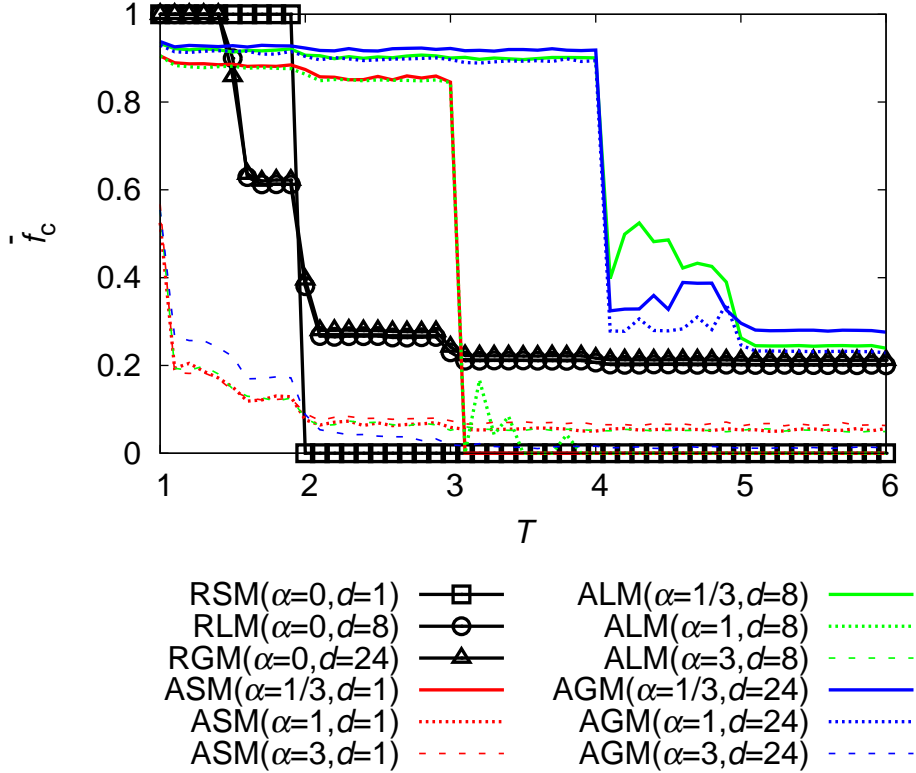}
\caption{Equilibrium fraction of cooperators ($\bar{f}_c$) as a function of temptation to defect ($T$). $\rho=0.3$.}
\label{fracC}
\end{figure}

\begin{figure*}[!t]
\centering
\includegraphics[width=6in]{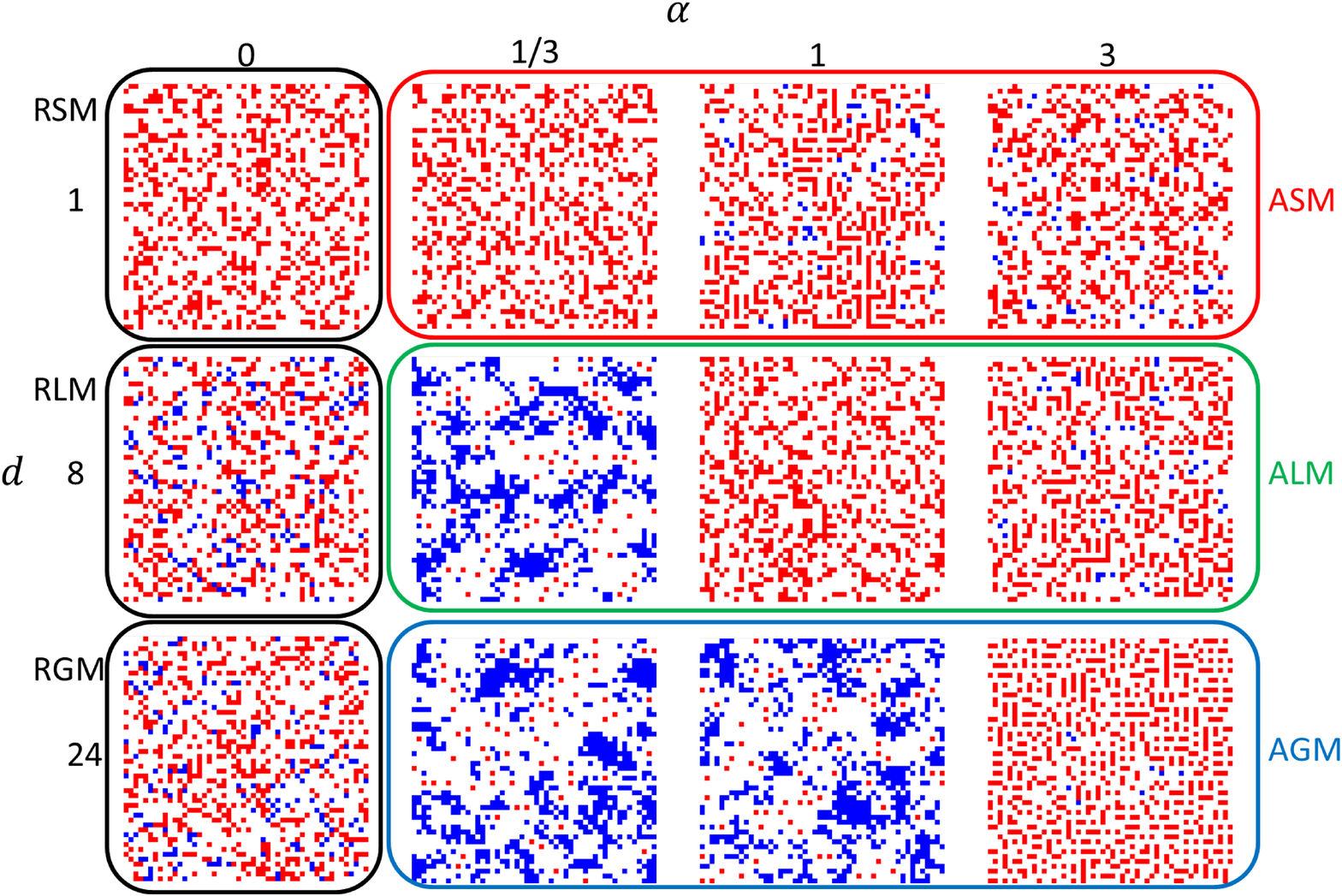}
\caption{Typical snapshots of simulations after 10,000 generations. Blue and red dots represent cooperators and defectors, respectively. Cooperators dominated in the population in ALM (Adaptive Long-range Migration; $\alpha=1/3$) and AGM (Adaptive Global Migration; $\alpha=1/3, 1$). In those cases, cooperators successfully maintained a certain distance with defectors. $T=3.5$ and $\rho=0.3$.}
\label{snapshots}
\end{figure*}

First and foremost, we found that AXM strongly promoted cooperation compared to RXM, especially under conditions where the temptation to defect was considerably high.
Figure \ref{fracC} shows $\bar{f}_c$ as a function of $T$. For the low values of $T$, all migration types
sustained a high level of cooperation (except for $\alpha=3$ that made migration rather infrequent).
For $T=1.5$, ALM and AGM ($\alpha=1/3$) were slightly lower (about 0.92) than RSM (full cooperation; 1.0) because individuals did not move any more after the complete segregation between cooperators and defectors (see Fig.~\ref{snapshots}).
In highly tempting conditions ($T>3$), however, cooperators were easily exploited by defectors in RXMs. For $T=3.5$, $\bar{f}_c$ was 0.00 in RSM. In ASM ($\alpha=1/3$), cooperators were allowed to move in response to the presence of defectors, but they often remained near the defectors and were exploited by them.
Therefore, cooperators were dominant up to $T=3.0$, but no longer survived when $T>3.0$.
In ALM and AGM, in contrast, cooperators were able to secure a certain distance from defectors after the adaptive migration, and thus cooperation was promoted even under highly tempting conditions.
For instance, for $T=3.5$, which was seemed to be extremely favorable for defectors, cooperators
were still dominant in the cases of ALM (0.90) and AGM (0.92) for $\alpha=1/3$.
Even complete segregation between cooperators and defectors was possible in ALM (for $\alpha=1/3$) and AGM (for $\alpha=1/3, 1$) (Fig.~\ref{snapshots}) (see also the typical movies of RXMs (Suppl.~Video S1, S2, and S3) and AXMs ($\alpha=1/3$) (Suppl.~Video S4, S5, and S6)).
Moreover, under high tempting conditions, $\alpha=1/3$ only led to high cooperation in the adaptive
long-range migration. This means that sensitivity to defectors is important in such situations (see Suppl.~Fig.~S1 for the effect of the $\alpha$-$d$).

\subsection*{Effect of density}
\begin{figure*}
\includegraphics[width=1\textwidth]{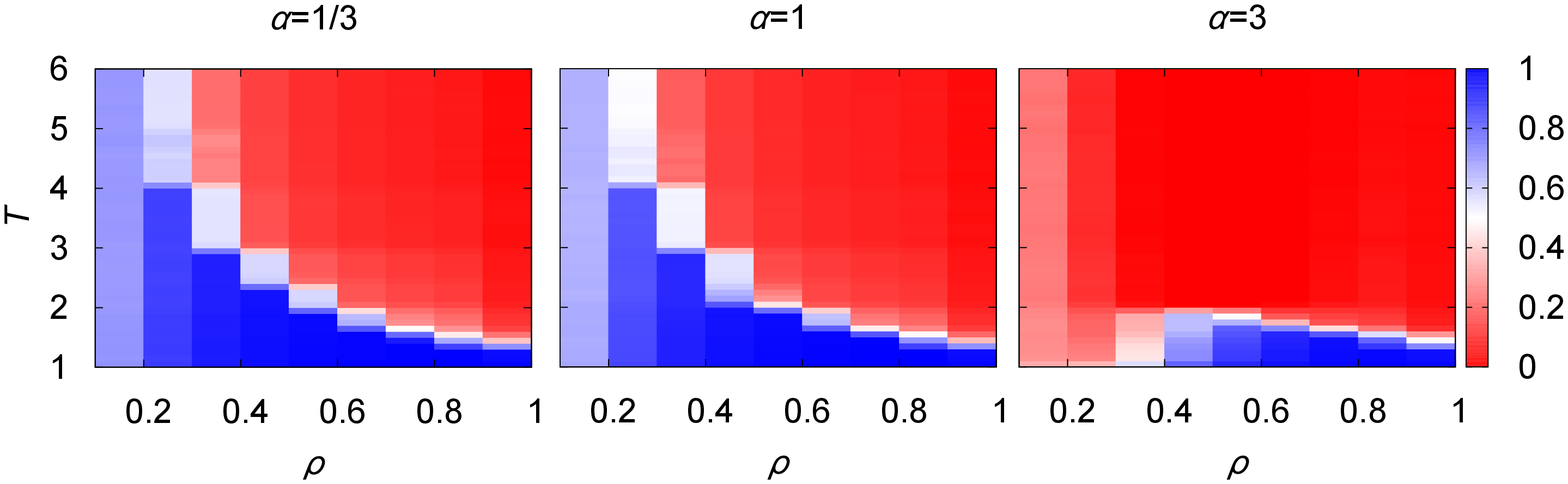}
\caption{Equilibrium fraction of cooperators ($\bar{f}_c$) as functions of population density ($\rho$) and temptation to defect ($T$). AGM (Adaptive Global Migration) was used as the mode of adaptive migration.
The results are obtained by averaging $f_c(t)$ for $t=4,001\sim5,000$ (averaged over 10 independent simulation runs).}
 \label{density}
\end{figure*}

We also investigated the effect of the population density on the evolution of cooperation.
Figure \ref{density} summarizes the results of another set of systematic simulations with varying $\rho$ and $T$. AGM was used as the mode of migration.
It can be seen in these plots that there is an optimal population density (e.g., around $\rho = 0.3$ for $\alpha=1/3$) that best promotes cooperation. 
If the density is extremely low, cooperators cannot form clusters, while if the density is too high,
cooperators will likely meet other defectors even after adaptive migration.
Our finding, that the moderate densities are suitable for enhancing cooperation, is in agreement with results reported in earlier literature.

\subsection*{Robustness against noise}
We also tested the robustness of the results against noise, i.e., stochastic changes of individuals' strategies.
This is an important issue because, without such noise, it would be rather trivial that the clusters of cooperators can sustain themselves forever once they are established separately from defectors.
Realistically, however, such clusters would be susceptible to defectors emerging from inside, if strategic changes are allowed.
To assess the impact of such strategic changes on the evolutionary sustainability of cooperation,
we implemented random mutations of individual strategies.
More specifically, after an individual imitates the highest strategy from its neighbors, the imitated strategy may flip to the opposite with probability $p_m \ll 1$.

\begin{figure}[!t]
\centering
\includegraphics[width=5in]{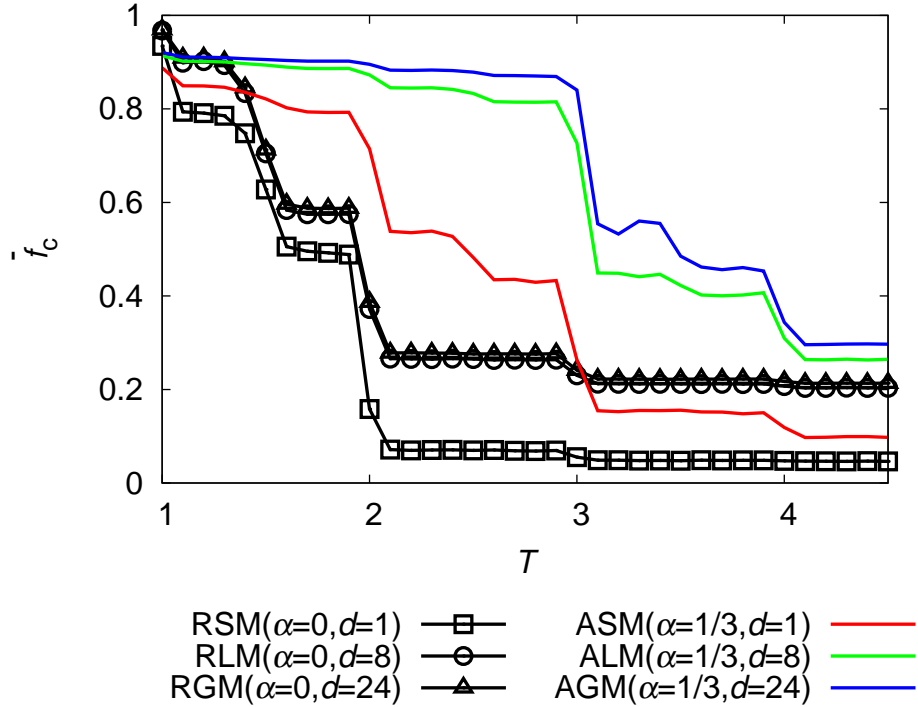}
\caption{Equilibrium fraction of cooperators ($\bar{f}_c$) as a function of the temptation to defect ($T$) with mutation. $p_m=0.01$.}
\label{mut_fracC}
\end{figure}

Supplemental Fig.~S2 shows typical evolutionary simulation runs with and without mutation ($\alpha=1/3$).
Compared to no-mutation cases, conditions with mutation generally resulted in a lower level of cooperators because mutations often destroyed cooperative clusters, causing the continuously fluctuating $f_c$.
However, the impact of mutation was much less for long-range migration cases (ALM and AGM) than for ASM.
Moreover, the sensitivity analysis over $T$ revealed that ALM and AGM still maintained high levels of cooperation compared to other conditions (Fig.~\ref{mut_fracC}).
These results confirmed that cooperation is promoted in ALM and AGM even under noisy conditions with strategic mutations.

\subsection*{Emergence of cooperation}
Lastly, we considered the hardest situation for cooperation in which the initial condition is made of 100\% defectors.
In this setting, cooperators can still appear by mutation, but always as a minority surrounded by defectors.
Supplemental Fig.~S3 shows typical simulation runs with mutation starting from 100\% defector initial populations.
It is observed that cooperation did emerge by mutation, and grew and maintained a certain level of proliferation (similar to those see in Suppl.~Fig.~S2) throughout the simulation runs.

\begin{figure}[!t]
\centering
\includegraphics[width=5in]{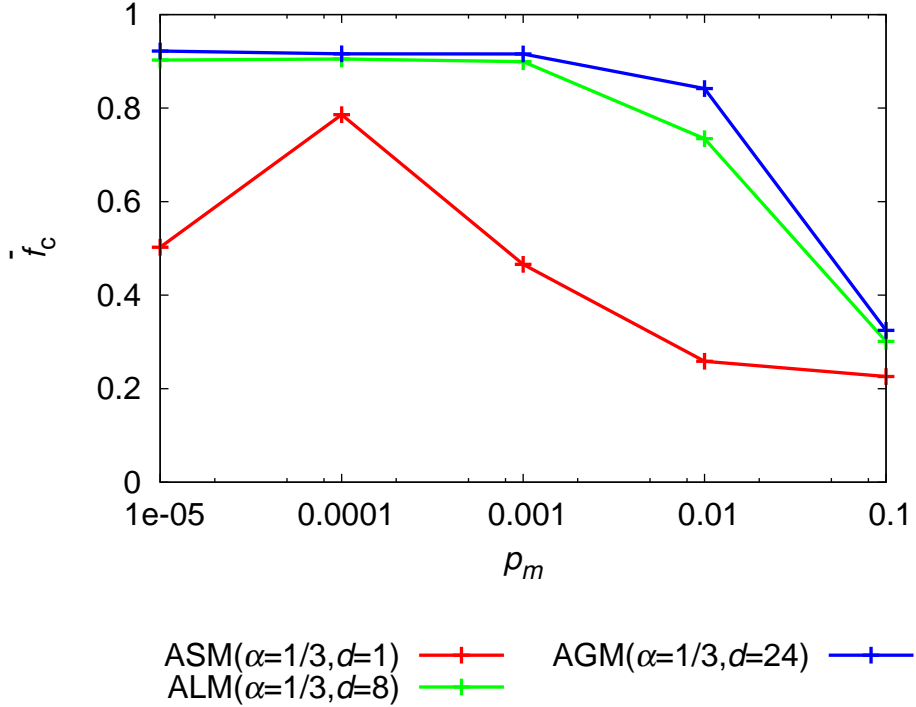}
\caption{Robustness of cooperation against the mutation probability ($p_m$). $\alpha=1/3$, $T=3.0$, and $\rho=0.3$.}
\label{mut_robust}
\end{figure}

We then conducted a sensitivity analysis of this result over varying mutation rate $p_m$ (Fig.~\ref{mut_robust}).
To our surprise, cooperation successfully emerged and dominated the population for extremely low mutation rates ($p_m<10^{-3}$) especially in ALM and AGM (Fig.~\ref{mut_robust}).
We note that there are at least two important factors for understanding why cooperation can emerge and fixate with such low mutation rates. First, in our model setting, migration occurs before game play and strategy updating, so a mutant cooperator that emerged in a defector-dominated population in the previous generation may still be able to escape
from the nearby defectors before strategy updating, if adaptive long-range migration is allowed. If the strategy updating were to take place before migration, 
the new cooperator would quickly become reverted back to a defector that is more successful in terms of payoffs.
Second, our model assumed that strategic mutation would occur only immediately after an imitation, so a mutant cooperator can maintain the same cooperative strategy indefinitely as long as there are no neighbors around it.
Due to this, mutant cooperators that luckily found an isolated empty space due to long-range migration can survive, and once such cooperators come next to each other, cooperative behavior can spread quickly by clustering.
Moreover, these clusters are less likely to be destroyed if the mutation rate is low.

\section*{Discussion}
In this paper, we investigated the effect of adaptive long-range migration on the evolution of cooperation in a spatial population model.
We found that adaptive long-range migration strongly enhanced the evolution and maintenance of cooperative behavior, especially under high tempting conditions where other forms of migration did not.
We also found that the cooperation emerging with the adaptive long-range migration was considerably robust against stochastic mutations of strategies and the initial lack of cooperators in the population.

A possible caveat against this study is the robustness of cooperation may depend on the timing
between migration and strategy updating. In other words, whether migration before strategy updating is appropriate or not.
Based on the previous study \cite{Jiang_etal2010}, we have assumed that migration occurs prior to strategy updating.
This point should be discussed more in future works.

In the present study, we have not assumed individuals' global knowledge about the environment, in contrast to the
previous studies such as Helbing et al.~\cite{HelbingYu2008, Helbing2009, HelbingYu2009} where individuals can know not only the
quality of the present location but also the desire destination within the neighbors to move.
For humans and animals, this situation may be hard to occur because new environments for individuals are often unpredictable in the real world.
Therefore, the assumption in our model that individuals escape from the bad conditions but they randomly seek new environments is plausible.

Moreover, adaptive long-range migration is a naturally observed in human and animal behaviors.
We can see that wild animals show a great migration when they fall into a starvation state.
In addition, humans can easily detect unfavorable situations and then sometimes migrate globally by means of transportation.
In those senses, adaptive long-range migration (but the destination is random) is appropriate for describing the migration of the species.

We believe that our model and findings provide a solid theoretical advancement for understanding the evolution of cooperation, which has still remained an open question in various scientific fields.
Compared to earlier studies that assumed only short-range migration, our work illustrated that cooperative behavior may arise and proliferate more easily if one takes into account organisms' adaptive responses that scale with the strength of external stimuli --- a typical behavioral trait commonly seen in animals and humans.

\section*{Methods}
We develop an agent-based model in which individual agents distribute over a two-dimensional square lattice and 
play the PD game with their neighbors.
The square lattice is composed of $L \times L$ sites with periodic boundary conditions. Each site is either empty or occupied by one individual.
Empty sites represent spatial regions that individuals can migrate to.
Initially, individuals are randomly distributed over the square lattice. One half of them are set to be cooperators while the other defectors.
Their population density is given by $\rho$ (i.e., the fraction of empty sites is $1-\rho$). The population density remains constant throughout a simulation run, since individuals will never die or be born.

Individuals are updated asynchronously in a randomized sequential order. The algorithm for updating an individual consists of the following three phases:
\begin{enumerate}
\item {\em Migration.} 
Extending the adaptive migration model proposed by Jiang et al.~\cite{Jiang_etal2010}, our model incorporates adaptive long-range migration as follows.
First, to decide the site to move into, the individual counts the number of defectors, $n_D$, within its $n$ neighbors.
Using $n_D$, the maximum migration distance is calculated as $d_{\mathrm{max}}=\mathrm{round}((n_D/n)^{\alpha}d)$, where $\alpha$ and $d$ are experimental parameters.
We call $\alpha$ sensitivity to defectors and $d$ migration range, respectively.
$\mathrm{round}(...)$ is used here because the space is discrete (see Fig.~\ref{d_max}). 
Then the individual moves to a randomly selected empty site within a $(2d_{\mathrm{max}}+1)$-by-$(2d_{\mathrm{max}}+1)$ square region around its original location (see the example in Fig.~\ref{d_max_sample}).
This ``adaptive long-range migration" enables individuals to migrate to a more distant site when the number of nearby defectors is larger.
If there is no empty site within the region, the individual remains in the original location.
\item {\em Game play.}
After the migration, the individual plays the PD game with other individuals within its $n$ nearest neighbor sites and accumulates the payoffs resulting from the games.
If there are no other individuals within the neighborhood, no game is played.
In each game, two individuals decide whether to cooperate or defect simultaneously based on their current strategies.
They both obtain payoff $R$ for mutual cooperation while $P$ for mutual defection.
If one selects cooperation while the other does defection, the former gets the sucker's payoff $S$ 
while the latter gets the highest payoff $T$, the temptation to defect.
The relationship of the four payoffs is usually $T>R>P>S$ in PD games.
Following the parameter settings used in the model by Nowak and May \cite{NowakMay1992}, we used $P=0, R=1$, and $S=0$, while $T>1$, the temptation to defect, was varied as an experimental parameter.
\item {\em Strategy updating.}
Once all the games are over, the individual imitates the strategy of the other individual that achieved the highest total payoff among its neighbors (including itself; if there is a tie one individual is randomly selected).
If there are no other individuals within the neighborhood, the individual maintains the same strategy.
\end{enumerate}

When all individuals finish conducting the three phases above, it constitutes one time step of the simulation, which we refer to as a ``generation'' hereafter.

\providecommand{\noopsort}[1]{}\providecommand{\singleletter}[1]{#1}%

\section*{Author contributions}
GI and HS designed the research. GI and MS constructed the model. GI performed the simulation.
GI, HS, and DSW discussed and analyzed the results. GI and HS wrote the main manuscript text.
GI, HS, and DSW reviewed the manuscript.

\section*{Additional information}
\subsection*{Competing financial interests}
The authors declare no competing financial interests.

\section*{Figure legends}
Figure 1. Maximum migration distance $d_{\mathrm{max}}$ as a function of the number of defectors in the neighborhood ($n_D$). The Moore neighborhood is assumed (i.e., $n=8$). Left: Short-range migration ($d=1$). Center: Long-range migration ($d=8$). Right: Global migration ($d=24 \sim L/2$). Random migration models do not depend on $n_D$ (red lines at the top), while adaptive migration models (all the other curves) depend on $n_D$.\\

Figure 2. An illustration of how ALM (Adaptive Long-range Migration) works ($\alpha=1$, $d=8$). In this example, there are three defectors within eight neighbors indicated by the red square (i.e., $n=8$, $n_D=3$). The maximum migration distance is calculated as follows: $d_{\mathrm{max}}=\mathrm{round}((3/8)^1 \times 8)=3$. Therefore, the focal cooperator can migrate to another empty site within the 7-by-7 local region indicated in green.\\

Figure 3. Equilibrium fraction of cooperators ($\bar{f}_c$) as a function of temptation to defect ($T$). $\rho=0.3$.\\

Figure 4. Typical snapshots of simulations after 10,000 generations. Blue and red dots represent cooperators and defectors, respectively. Cooperators dominated in the population in ALM (Adaptive Long-range Migration; $\alpha=1/3$) and AGM (Adaptive Global Migration; $\alpha=1/3, 1$). In those cases, cooperators successfully maintained a certain distance with defectors. $T=3.5$ and $\rho=0.3$.\\

Figure 5. Equilibrium fraction of cooperators ($\bar{f}_c$) as functions of population density ($\rho$) and temptation to defect ($T$). AGM (Adaptive Global Migration) was used as the mode of adaptive migration.
The results are obtained by averaging $f_c(t)$ for $t=4,001\sim5,000$ (averaged over 10 independent simulation runs).\\

Figure 6. Equilibrium fraction of cooperators ($\bar{f}_c$) as a function of the temptation to defect ($T$) with mutation. $p_m=0.01$.\\

Figure 7. Robustness of cooperation against the mutation probability ($p_m$). $\alpha=1/3$, $T=3.0$, and $\rho=0.3$.

\end{document}